# Median Algorithm for Sector Spectra Calculation from Images Registered by the Spectral Airglow Temperature Imager

*Atanas Marinov Atanassov*

*Space and Solar-Terrestrial Research Institute, Bulgarian Academy of Sciences,
Stara Zagora Department, Bulgaria,  At_M_Atanassov@yahoo.com*

*The Spectral Airglow Temperature Imager is an instrument, specially designed for investigation of the wave processes in the Mesosphere-Lower Thermosphere. In order to determine the kinematic parameters of a wave, the values of a physical quantity in different space points and their changes in the time should be known. As a result of the possibilities of the SATI instrument for space scanning, different parts of the images (sectors of spectrograms) correspond to the respective mesopause areas (where the radiation is generated).*

*An approach is proposed for sector spectra determination from SATI images based on ordered statistics instead of meaning. Comparative results are shown.*

**Introduction**

Spectral Airglow Temperature Imager is optical space scanning instrument developed in Centre for Research in Earth and Space Science (CRESS) by York University Canada [1, 2]. For the oxygen molecule emissions, corresponding to transitions $O_2(b^1\Sigma_g^+ - X^3\Sigma_g^-)$ the maximum radiation is at the height of the mesopause. It is possible to determine the temperature and the emission rate at the height of the mesopause in different points of an annular sky segment with centre in the zenith by registration of this emission with the SATI instrument and applying suitable data processing. This enables the investigation of the gravity waves propagated at the height of the emission radiation maximum. Except for gravity waves [3, 4], SATI data are used for investigation of planetary waves [5], as well as for the seasonal course of some parameters characterizing the dynamics of the Mesosphere/Low Thermosphere [6].

A version SATI-3SZ was developed at the Stara Zagora Division of the Solar-Terrestrial Influences Laboratory in collaboration with CRESS Laboratory [7].

A new algorithms for data processing registered with this SATI instrument [8÷11] are developed. This paper presents a new approach for calculation of sector spectra. Some comparative results are also presented.

**Sector spectra calculation by classical algorithms for SATI image processing**

In original algorithms [3, 4] of SATI data processing the images (Fig. 1a) are divided on 12 sectors with equal angle by 30° each (Fig.1b). This is possible after some preliminary image processing connected with cosmic ray rejection and dark image correction and determination of the image centre coordinates $(i_0, j_0)$. These coordinates are calculated with precision of a whole pixel along each of the two directions of the axes Ox and Oy. Every pixel at a distance less than the image radius (~128ps) is checked in which sector falls according to the angle between its radius vector and the basic direction of the axis Ox

$$\gamma = a\tan((i - i_0)/(j - j_0)),$$

Besides, according to the distance to the image centre p (in pixels), the values of the registered pixel intensity is summed up with the intensities of all pixels located at the same distance:
if $\gamma \in (\gamma_{1,k}, \gamma_{2,k})$ , then

$$S_k(p) = S_k(p) + I(i,j) , \ p = \sqrt{(i-i_0)^2 + (j-j_0)^2} ,$$

where $S_k$ is $k^{-th}$ sector sum in the pixel space and p accepts the rounded value of r. Finally, all produced sums for the sector are averaged in accordance with the number of pixels at the same distance from the centre; in the same way the value of a spectrum element (Fig. 1c) is determined for the respective sector

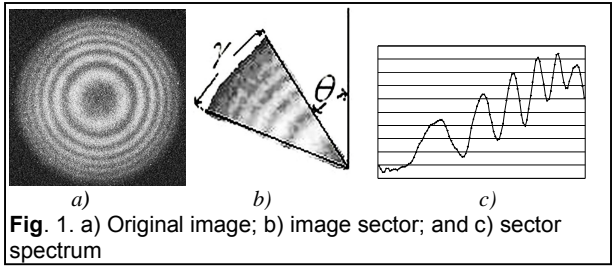

*a) b) c)*
**Fig**. 1. a) Original image; b) image sector; and c) sector spectrum

$$S_p^k = S'^k_p / n_p^k, \quad n_p^k = \sum_{p=\sqrt{(i-i_0)^2+(j-j_0)^2}} I_{i,j}$$

The spectra measured by the SATI instrument are not linear in relation to the wavelength [7]. The spectra are presented as series of intensities as a function of the distance to the image centre. The correspondence between the serial number of each pixel and the wavelength is significant for the interpretation of the registered spectra, not for their extraction from the images. Thus, each spectrum is formed as a series of values of registered intensities for the respective wavelength, corresponding to the respective pixel.

Each sector spectrum is compared with preliminary calculated spectra (synthetic spectra) for different temperatures of the radiating gas (in our case rotation spectra of $O_2$), convoluted with the transmittance function of the employed interference filter.

**Previous development of sector spectre determination algorithm**

A flexible approach for sector spectra determination is presented in [11], with which the sector parameters ($\gamma,\theta$) are selected arbitrarily within certain intervals. The angle $\theta$ varies in the interval (0÷360°) with a step $\Delta\theta$, for example ~1°. The angle $\gamma$ is selected arbitrarily from some degrees to some dozens of degrees, depending on the space noise in the image. The overlapping of the adjacent sectors is obviously due to this approach.

The distance to the image centre is determined for all pixels in a given sector. The pixel intensities, located at an equal distance from the centre, determined with precision up to one pixel, are summed up and averaged according to their number for each distance. In this way averaged sector spectra are obtained.

Unlike the original approach, here the image centre is determined with precision, higher than one pixel [8], thus enabling a more precise determination of the distances from the image centre to the sector pixels. In this way the sector spectra are determined more precisely. This is evidenced by the more precise determination of the sector temperatures.

**New approach for sector spectra determination**

Instead of calculating the values of the measured sector spectra as mean values of the measured intensities for all pixels in a particular sector, located at an equal distance towards the image centre, another approach was developed and tried. Again the values of all pixels located at an equal distance p towards the image center are used for determination of one value of the sector spectrum. The verification of all pixels of the sector is organized by an algorithm, identical to the one, described in [11].

**Version I:** the sector is entirely in one of quadrants I, II, III or IV only and we can write down for index by Ox axis

$$i \in (\text{NINT}(\sin g(1,\cos\varphi_1)), \text{NINT}(R * \text{MAX}(\cos\varphi_1, \cos\varphi_2))),$$

for $\cos\varphi_1 > 0$

and

$$i \in (\text{NINT}(\sin g(1,\cos\varphi_1)), \text{NINT}(R * \text{MIN}(\cos\varphi_1, \cos\varphi_2))),$$

for $\cos\varphi_1 < 0$

**Version II:** the sector falls into I and IV or II and III quadrants simultaneously. Then for the changing of the index by Ox axis

$$i \in (\text{NINT}(\text{sign}(1,\cos\varphi_1)), \text{NINT}(\text{sign}(R,\cos\varphi_2))) .$$

We will note only that in the versions described here, $\text{sign}(\cos(\varphi_1)) = \text{sign}(\cos(\varphi_2))$. The function NINT() is intrinsic function in the Fortran programming language which returns the nearest integer to the argument [12]. The functions MAX() and MIN() return the maximum or minimum value respectively of the arguments.

For the change of index by Oy axis in the above two versions we can write down

$$j \in (\text{NINT}(k_1.i), \text{NINT}(k_2.i)), k_1.i < k_2.i$$

or

$$j \in (\text{NINT}(k_2.i), \text{NINT}(k_1.i)), k_2.i < k_1.i$$

**Version III:** the sector falls into I and II or III and IV quadrants simultaneously. Then

$$i \in (\text{NINT}(R.\cos\varphi_1), \text{NINT}(R.\cos\varphi_2)), \cos\varphi_1 < \cos\varphi_2$$

or

$$i \in (NINT(R.\cos\varphi_2), NINT(R.\cos\varphi_1)), \cos\varphi_1 > \cos\varphi_2$$

For the index change along the Oy axis in the last third version we have for $\cos\varphi_1 < \cos\varphi_2$

for $i \in (NINT(R.\cos\varphi_1), 0)$   $j \in (k_1.i, \sqrt{R^2 - i^2})$

and for $i \in [0, NINT(R.\cos\varphi_2))$   $j \in (k_2.i, \sqrt{R^2 - i^2})$

Analogously, the boundaries in which the indices change and by $\cos\varphi_1 > \cos\varphi_2$ can be determined.

Figure 2 illustrate above described versions.

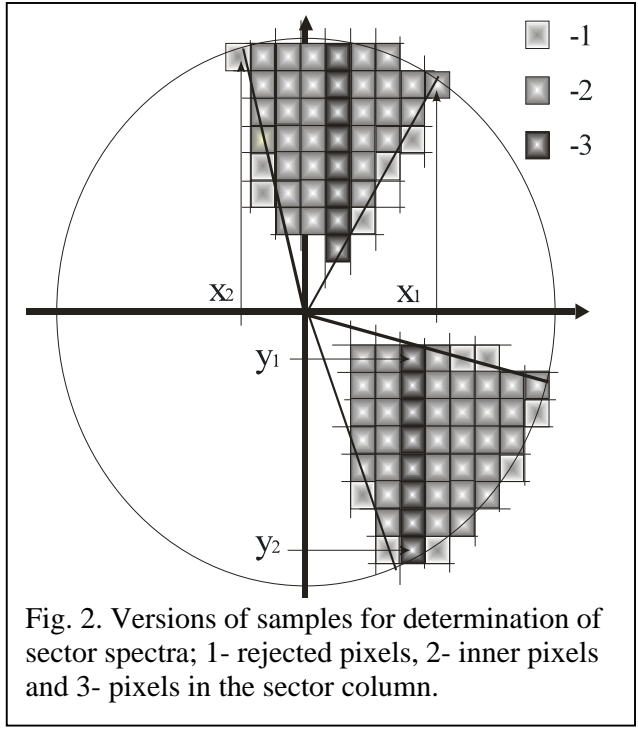

Fig. 2. Versions of samples for determination of sector spectra; 1- rejected pixels, 2- inner pixels and 3- pixels in the sector column.

Simultaneously with the determination of the distance for each pixel towards the image center, their values are stored in rows of a 2-dimensional area $W_{q,p}$. Each column p of the area contains the read values for all pixels located at distance p towards the image center. The sizes of the area are determined so as all measured values of the pixels in the sector with radius R(~125p) and angle γ (5÷40°) to be stored. The application of a bi-dimensional structure presumes a non-efficient utilization of the storage. However, this approach is suitable because the number of pixels is small, additional pointers are not used and the realization is simple, and the sorting is effectively applied. Tentatively, dimension q of area W is determined as

$$q = \gamma.R + \Delta,$$

where Δ is an additional quantity.

The standard run-time subroutine **sortqq** from the Visual Digital Fortran library is used to sort the values in the W columns.

**Some results from the application of the algorithm**

a) *differences between sector spectra temperatures determined by different approaches*; sector spectra temperatures and standard deviations retrieved by ordinary (s) and median (m) approaches are shown in Figure 3 a,b and Figure 4 a,b. It is obvious that the proposed approach works since the differences between the two approaches are very small;

b) *sector spectra dependence on the sector angle*; the dispersion of the nocturnal courses of the errors when determining the temperatures as well as the sector temperatures decreases by increasing the sector angle (Fig. 4a,b);

c) *differences by the produced temperatures with the two approaches;* when increasing the sector angle (the filter window), the dispersion in the nocturnal course of the determined temperature decreases (Fig. 5a,b).

The differences between the two approaches are within the range of some degrees at sector angle of 5deg and get closer when it is increased.

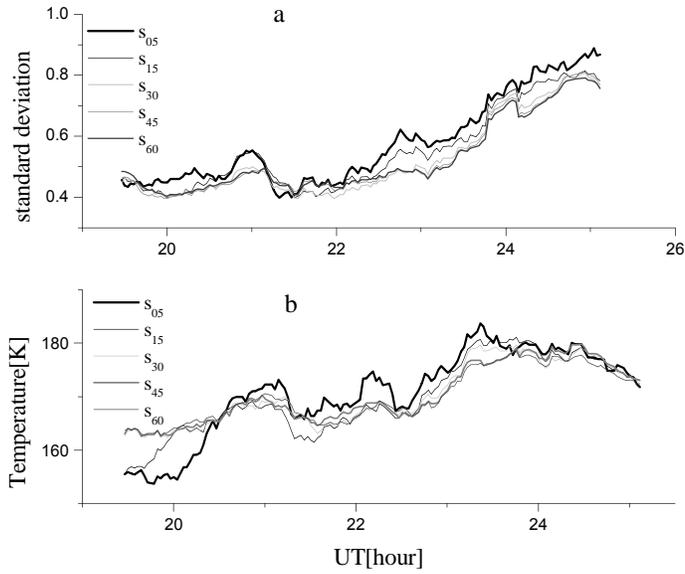

**Figure 3.** Denoised nocturnal courses of standard deviation (a) and temperature (b) for sectors with angles 5, 15, 30, 45 and 60deg calculated by median approach.

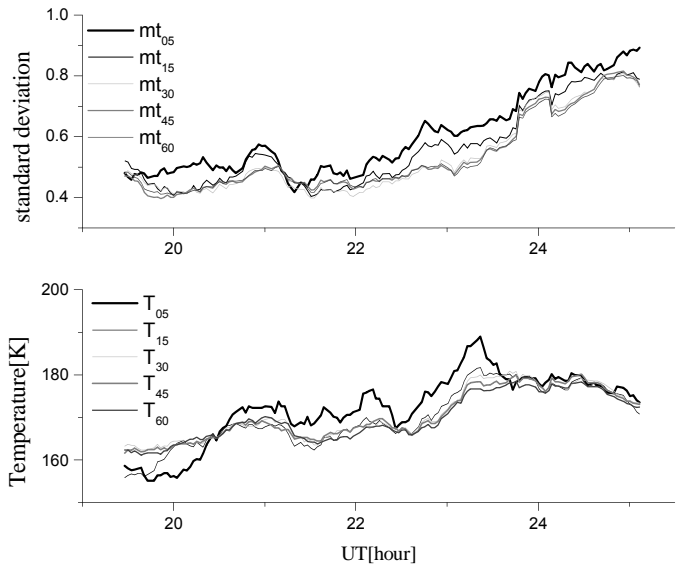

**Figure 4.** Denoised nocturnal courses of standard deviation (a) and temperature (b) determined by median approach for sector with angle 5, 15, 30, 45 and 60deg;

## Conclusion and future work

Unlike the previous algorithms for sector spectra determination where the intensities were calculated by averaging, now the values of all pixels at an equal distance towards the image centre are sorted. The value of the middle element in the ordered area after the sorting is taken as the intensity value, analogously to the median filtering. That is why the pixel values which contain pulse noise (salt and pepper noise- results from high energy particles) are not involved in the determination of the intensity of the measured spectrum for the respective wavelength. The application of the sorting procedure is connected with more processing time than the averaging and depends on the number of pixels, i.e. on the sector angle. Different sorting methods exist [13] which yield equivalent results. The efficiency of the sorting methods is different and depends on the number of values for sorting. When this number is small, as in our case, the application of simple sorting methods is possible.

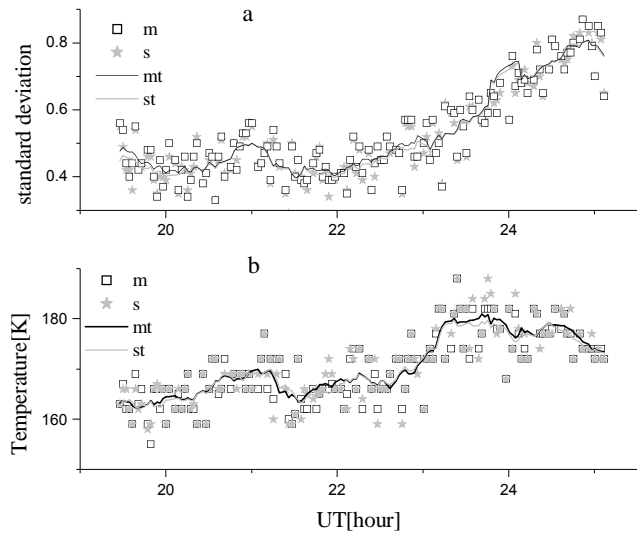

**Figure 5.** Nocturnal courses of standard deviation (a) and Temperature (b) determined for sector with angle 30deg; the symbol denotes the values calculated by "median" approach and the symbol ξ- by standard one by averaging. The lines "__" for "median" approach and "__" for standard one by averaging represent denoised series.

A space filtering is possible if the pixels located at distance (p-1) and (p+1) from the image centre and adjacent to those at distance p are sorted together for determination of the intensity of the $p^{-th}$ values in the spectrum. Let's note that the sorting procedure is non-linear and the mentioned possibility is not equivalent to its sequential applying on the pixels, located at distance p for determination of the sector spectrum and after that for the one-dimension filtering of the produced spectrum.

The application of other approaches for sector spectra calculation is possible. The approach, proposed here, is a special case of the order-statistic filters [14] application. An additional investigation would reveal the effects of the different approaches for sector spectra calculation and their significance for the sector temperature determination, which is the ultimate aim of the image processing.

**Acknowledgments:** The author would like to thank Dr. M.G. Shepherd for the support and encouragement and Mrs. Kr. Takucheva for the technical assistance.